\documentclass[aps,floatfix,longbibliography,notitlepage,twocolumn]{revtex4-2}
\usepackage{graphicx}
\usepackage{enumerate}
\usepackage{amsmath}
\usepackage[caption=false]{subfig}
\usepackage{hyperref}
\usepackage{float}
\usepackage{ulem}
\usepackage{xcolor}
\usepackage{soul}

\date{\today}

\begin{document}

\title{Analysis of spin avalanches due to interplay of disorder and temperature }
\author{Niharika Bhuyan} 
\author{Diana Thongjaomayum}
\email{dianat@tezu.ernet.in}
\affiliation{Department of Physics, Tezpur University, Assam 784028, India}

\begin{abstract}
The nonequilibrium zero-temperature Random Field Ising Model (RFIM) has been extensively studied to understand
critical response and avalanches in disordered driven systems. The emergence of power-law behaviour is observed over a wide region around the critical point. These studies however, are confined to zero-temperature dynamics. We study the role of temperature, which is inevitable in real experiments, in the context of RFIM on triangular lattices. We explore the interplay of different parameters: temperature, random field strength, and relaxation time which affect the prevalence of power-law behaviour on the lattice. The results indicate that power-law survives only in the regime of low temperature or small and intermediate disorder. Variations in temperature and disorder have similar affects on the avalanche-size distribution, indicating their strong correspondence. We also discuss the process of blurring out of the power law on increasing temperature or disorder. 

\end{abstract}
\maketitle
\section{Introduction}

Power-law distributions are one of the most common patterns found in nature. These patterns govern diverse phenomena such as Barkhausen noise from magnetic domain walls\cite{xu2015barkhausen,Bohn2018}, acoustic emissions during paper crumpling\cite{Acoustic}, the movement of tectonic plates during earthquakes\cite{Meng2019PowerLawxu}, etc. The beauty of power law lies in its universality, as the same mathematical rule extends from microscopic magnetic domain wall fluctuations to macroscopic phenomena such as earthquakes. This behaviour in several disordered systems can be explained by a simple model called the zero-temperature Random Field Ising Model (RFIM)\cite{sethna1993hysteresis,nattermann1998theory, belanger1991random}. 
  It is characterized by an interacting Ising spin system with a quenched disorder. The zero-temperature approximation is justified when the energy barrier to transition from one metastable state to another in the disorder-induced rugged energy landscape is very high\cite{ sethna2004random}  compared to the available thermal energy. Hence, thermal fluctuations can be neglected, when spin flips are solely driven by the external field to overcome the barriers. 
  Previous studies on zero-temperature RFIM reveal the presence of macroscopic jumps in the hysteresis curve as it approaches the critical point\cite{0temprfmi, fishman1979random, theodorakis2015random, dhar1997zero, avalanche, dferro, janicevic2021scaling}. Consequently, the spin avalanche distribution exhibits power-law behaviour around the critical regime.

Experiment on $2D$ semiconductor nanocrystal-FET array, composed of $InAs/CdSe/ZnSe$ nanocrystals, matched the avalanche dynamics of theoretical results given by RFIM\cite{Zohar2013}. In a recent experiment, researchers fabricated artificial spin ice arrays using nanomagnets; an external magnetic field was applied quasi-statically to mimic zero-temperature dynamics. The measured avalanche size distribution followed power-law behaviour in intermediate disorder\cite{PhysRevResearch.5.013011}. Many such experimental studies such as magnetic X-ray scattering on $\mathrm{Mn}_{0.75}\mathrm{Zn}_{0.25}\mathrm{F}_{2}$\cite{hill1993magnetic}, Barkhausen noise in thin $Fe$ films\cite{ironexpava}, magnetic-field-induced avalanches in a square artificial spin ice array of interacting nanomagnets\cite{2021experimental}, etc validate  RFIM.

\vspace{2mm} 
\par However, most experiments are performed at finite temperature, where thermal fluctuations play a significant role. In this paper, we investigate whether the characteristic power-law scaling observed at zero-temperature remains persistent when thermal effects are included. By studying this, we aim to bridge the gap between idealized zero-temperature predictions and the behaviour observed in finite-temperature physical systems. This is not the first time finite temperature RFIM has been studied. Previous research has explored RFIM at finite temperature on square lattices\cite{finitetemp2d} and $3D$ models\cite{3drfim, miga2009three}, revealing that macroscopic jumps vanish at finite temperature. The power-law behaviour is still observed only for small values of temperature and in an intermediate disorder. Temperature affects the critical exponent of this power-law, the exponent decreases when temperature is increased at high temperatures, but the critical exponent remains nearly constant at low and intermediate temperature\cite{Metra2021TemperatureDependent, Thermalves}. We extend this study on a $2D$ triangular lattice to understand the behaviour of the spin avalanches due to the interplay of disorder and temperature. We also discuss the relevance of relaxation time to ensure that the system remains in the nonequilibrium region manifested through hysteretic response. The study of the $2D$ system is currently crucial in view of the plethora of applications low-dimensional systems hold in terms of advanced and smart materials for technological importance.

\section{The Model}

The Hamiltonian of RFIM on the triangular lattice is\cite{nattermann1998theory}:

\begin{equation}
H 
= - \sum_{\langle i,j \rangle} J_{nn} s_i s_j - \sum_i h_i s_i - H_{\text{ext}} \sum_i s_i
\label{eq:hamiltonian}
\end{equation}
\\Here $J_{nn}(>0)$ is the ferromagnetic nearest-neighbor interaction, ${\langle i,j \rangle}$ indicates that only the nearest-neighbor spins are considered. 
$s_i$ is the Ising spin at site $i$, with values $\pm1$, denoting the up and down state. The localized random field $h_i$, is sampled from a Gaussian-distribution with standard deviation $\sigma$. 
The external magnetic field $H_{\text{ext}}$ is varied from negative to positive saturation using discrete steps in a cyclic manner.
The system is considered in contact with a heat bath at temperature $T$. Starting with all spins $s_i=\pm1$, we study the evolution of the spins using Glauber dynamics in a cyclic $H_{ext}$ varied in discrete steps on a periodic triangular lattice of $N\times N$ sites.

The spin flip probability is given by:
\begin{equation}
P(s_i) = \frac{1}{1 + e^{2Ke_i}}
\label{eq:probability}
\end{equation}
Here, $K=J/k_BT$, we assume $J=k_B=1$ and temperature, $T=1/K$. $e_i$ is the effective local energy for spin $s_i$\cite{glauberpro, glauber}. We sweep through the lattice, evaluating the spin-flip probability at each site for $t$ Monte-Carlo cycles. In one Monte-Carlo cycle, every site in the lattice is visited once to check for any spin update. After analyzing the spin-flip probability at each site for $t$ times, we calculate the magnetization using: 
\begin{equation}
m(H_{\text{ext}}) = \frac{1}{L_x L_y} \sum_i s_i
\label{eq:,magnetization}
\end{equation}
where $L_x$ and $L_y$ denote the dimensions of the lattice. 
For each increment in $H_{\text{ext}}$, all spins are initialized to the down-state($-1$) and then the Glauber update rule is applied. The process is repeated till $H_{ext}$ becomes large enough to flip all spins to the up-state. On reversing  $H_{ext}$ from positive to negative saturation, a hysteresis loop is traced out,
which helps to understand the key magnetic properties of a material, such as remanence, coercivity etc. 
When spins flip collectively for a fixed $H_{\text{ext}}$, we refer to it as an avalanche. We measure the avalanche by calculating the number of spins flipped in a particular $H_{\text{ext}}$ and record its frequency of occurrence to obtain an avalanche size distribution averaged over different disorder configurations.

By examining the scaling behaviour of avalanche distribution, the range of parameters where the system exhibits power-law behaviour are identified. 
In the upcoming sections, we discuss how the key parameters affect the system dynamics by varying one parameter at a time and fixing the others. This helps to understand the effect of each parameter on the system. In section~\ref{subsec:Vmct}, we study how the relaxation time, $t$ affects the system properties by fixing disorder strength $\sigma$ and temperature $T$. In section ~\ref{subsec:VDTL}, we investigate the role of disorder with fixed $t$ and $T$. Finally, in Section ~\ref{subsec:VTTL}, we analyze how temperature influences the system when $t$ and $\sigma$ are set.

\section{Results}
\subsection{Varying Monte-Carlo Cycle, $t$}
\label{subsec:Vmct}
\begin{figure}[h!]
   \centering
    \includegraphics[width=1\columnwidth]{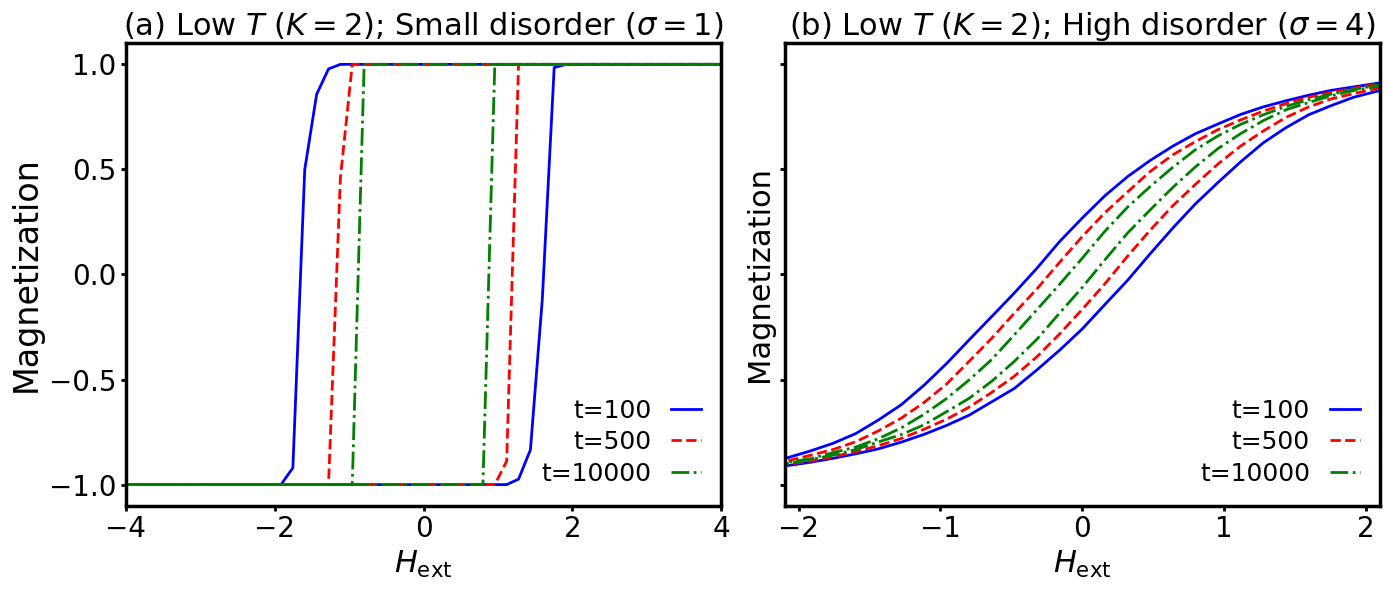}
  
\end{figure}
\begin{figure}[h!]
   \centering
    \includegraphics[width=1\columnwidth]{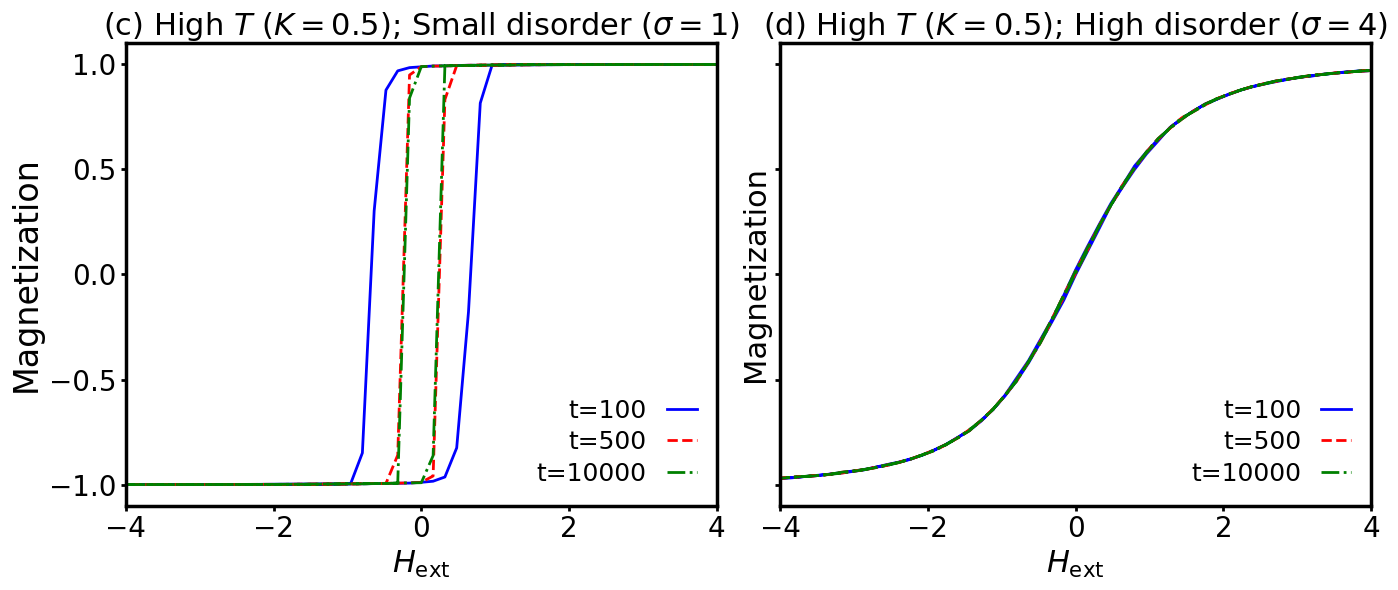}
    \caption{Hysteresis loops on $999\times 999
$ triangular lattice showing the effects of Monte-Carlo cycle, t= 100, 500, 1000, 10000 for different disorder $\sigma$ and temperature $T$($=1/K$) regimes: a)Small disorder ($\sigma=1$); Low T ($K=2$) b)Large disorder ($\sigma=4$); Low T ($K=2$) c)Small disorder ($\sigma=1$); High T ($K=0.5$) d)Large disorder ($\sigma=4$); High T ($K=0.5$).}
    \label{fig:disorderhyst}
\end{figure}

Figures $1a$ - $1d$ show hysteresis loops on a $999\times999$ triangular lattice for different Monte Carlo cycles $t$=100, 500, 1000 and 10000. The figures show different combinations of temperature ($K=2, 0.5$) and disorder ($\sigma=1, 4$) and their dependence on $t$. We observe that when $t$ is increased, the hysteresis loop width becomes narrower. This occurs because a large $t$ implies more relaxation time for the spins to approach equilibrium, which is evident from the disappearance of hysteresis. This effect is strongly dependent on temperature and disorder. At low temperature ($K=2$) with weak disorder ($\sigma=1$), as shown in Fig. 1(a), the loop width narrows significantly on increasing $t$. The narrowing rate decreases when disorder is increased to $\sigma=4$ for same $K=2$(Fig. $1(b)$). Similarly, comparing figures 1(a) and 1(c) reveals that raising the temperature from $K=2$ to $0.5$ at fixed disorder ($\sigma=1$) also reduces the narrowing rate. However, there is a limit to which the loop-width can be decreased. After a certain point, even on increasing $t$, the loop-width remains the same when either temperature or disorder is low (discussed below).
Fig. 1(d) shows a clear exception, where at high temperature ($K=0.5$) and strong disorder ($\sigma=4$), the system equilibrates rapidly within small $t$ and hence hysteresis is not observed for any values of $t$ considered here.

\begin{figure}[h!]
       \centering
        \includegraphics[width=1\linewidth]
    {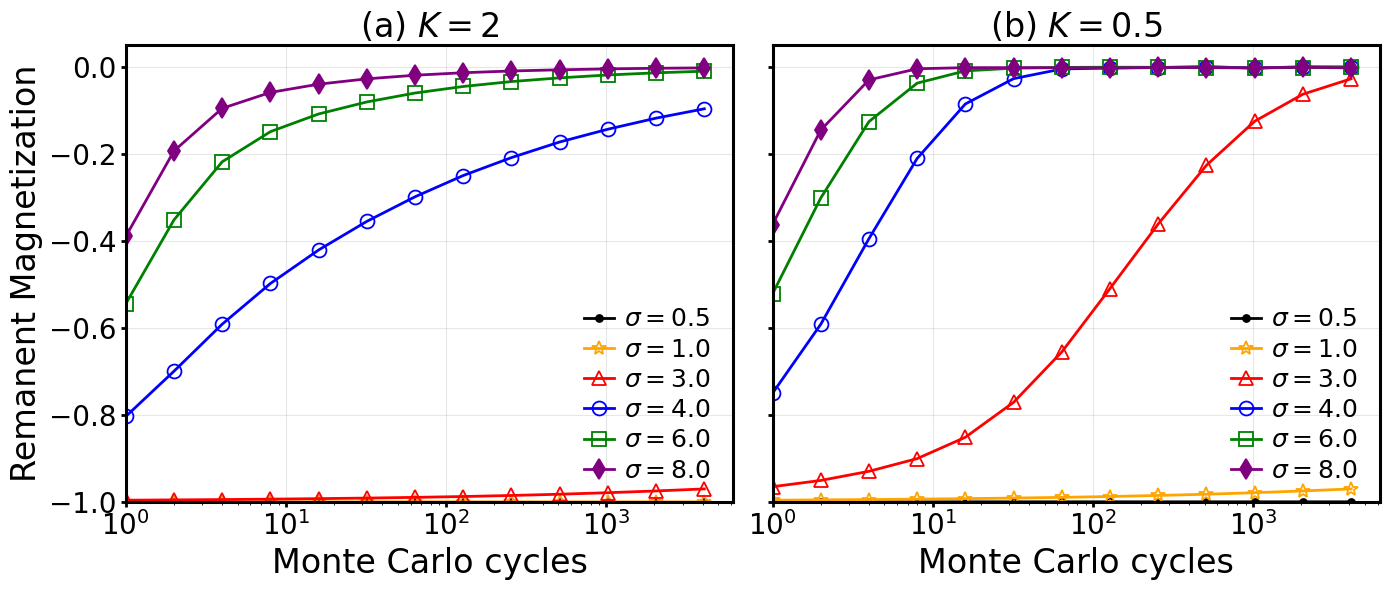}
     \caption{Disorder-dependent evolution of remanent magnetization versus Monte-Carlo cycles, $t$ for a $999 \times 999$ triangular lattice at: a)$K=2$ and b)$K=0.5$ for $\sigma=1,2,3,4,6,8$ averaged over 20 disorder realizations.}
    \label{fig:remamdisorder}
    \end{figure}

\begin{figure}[htp]
        \centering
        \includegraphics[width=1\linewidth]
    {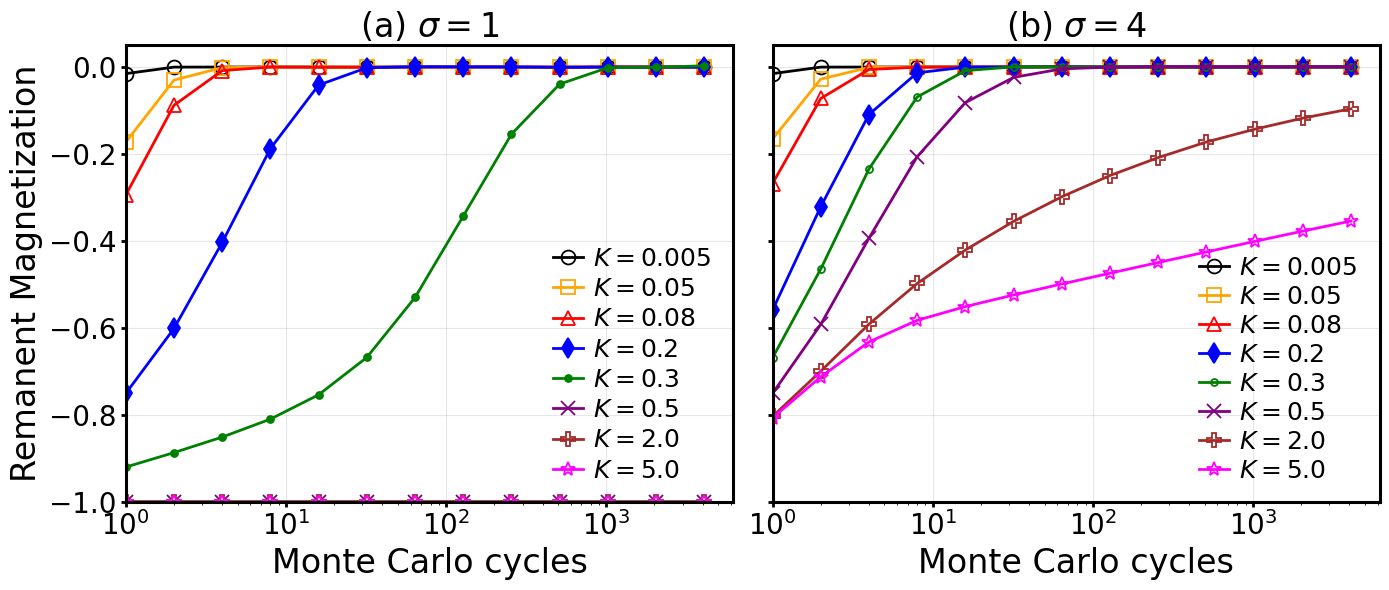}
   \caption{Remanence magnetization  vs $t$ for different temperatures $K=0.005,0.05,0.2,0.3,0.5,2,5$ on the lower hysteresis loop at: a)$\sigma=1$ and b)$\sigma=4$.}
    \label{fig:remtemp}
    \end{figure}

In addition to the hysteresis loop, the signature of the attainment of equilibrium state can be assessed through remanent magnetization $r(m)$, which is the magnetization retained by a material when \( H_{\text{ext}} \)$=0$. Thus, $r(m)=0$ implies that the system has reached equilibrium.
Fig 1b clearly shows how $r(m)$ decreases when $t$ is increased for the chosen parameters. 
More precisely, how the hysteresis loop closes and at what rate for a chosen parameter of $K$ and $\sigma$ is evident from the analysis of $r(m)$ vs $t$.

Fig.~\ref{fig:remamdisorder} shows the evolution of remanent magnetization on the lower hysteresis loop with Monte-Carlo cycles, $t$ at different disorder $\sigma= 0.5, 1, 3, 4, 6, 8$. 
Fig.~\ref{fig:remamdisorder}(a) corresponds to the low-temperature regime($K=2$) and Fig.~\ref{fig:remamdisorder}(b) represents the high-temperature regime($K=0.5$). 
All curves are averaged for 20 different configurations of localized random disorder on  $999\times999$ triangular lattice. 
The figures show that as $t$ increases, the remanent magnetization increases from a negative value and tends to saturate at zero rapidly for larger $\sigma$. 
However, at small disorder values ($\sigma=0.5, 1$), $r(m)$ remains frozen at $-1$ regardless of $t$, as at small disorder, the energy landscape remains relatively uniform until sufficient \( H_{\text{ext}}>0 \) is reached, and only very few spins could flip up. However, on approaching positive $H_{ext}$, a collective flipping occurs, evident from the magnetization loop for small $\sigma$.

Similarly, Fig.~\ref{fig:remtemp} illustrates the variation of $r(m)$ with $t$ at different temperatures for fixed: a) weak disorder ($\sigma=1$) and b) strong disorder ($\sigma=4$). It is observed that saturation occurs more rapidly at higher temperatures than at lower temperatures, irrespective of disorder. The remanent magnetization remains frozen at $-1$ only for the case of low temperature and low disorder, as spin flip probability remains low as evident from Eqn~\ref{eq:probability}. 
From Fig.~\ref{fig:remamdisorder} and Fig.~\ref{fig:remtemp}, we observe that high temperature and high disorder equilibrates the system faster.
This is indicated by the value of remanent magnetization approaching zero, and consequently the hysteresis loop tending to vanish, thereby indicating that the system has reached equilibrium.

We do not desire the equilibrium state as we want to study the power-law behaviour which exists in the non-equilibrium hysteretic region. 
 
Therefore, we select an appropriate range of $t$ from these graphs to ensure that the system remains in a non-equilibrium state for the chosen values of $K$ and $\sigma$ discussed in later sections. 

\subsection{Varying Disorder}
\label{subsec:VDTL}

\begin{figure}[h!]
      \centering
        \includegraphics[width=1\linewidth]
    {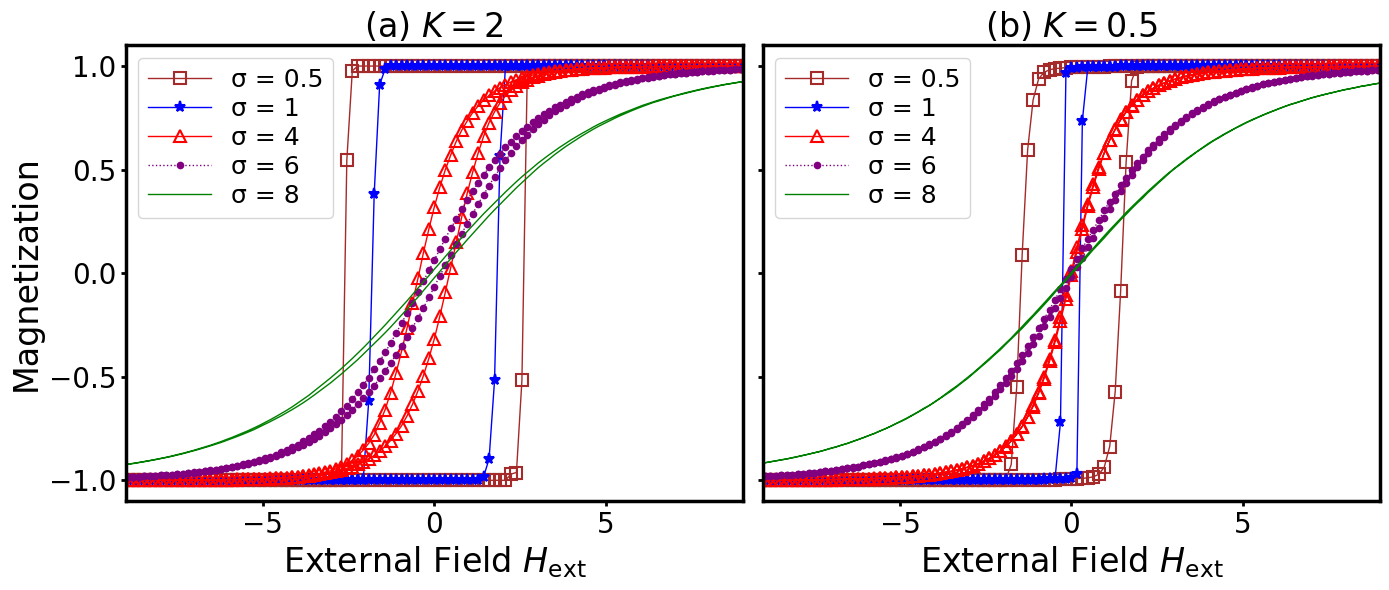}
    \caption{Hysteresis on a $999 \times999 $ triangular lattice for different disorders($\sigma=0.5,1,4,6,8$) at $t=50$ for fixed temperatures: a)$K=2$ and b)$K=0.5$.}
    \label{fig:disortrih}
    
\end{figure}

\begin{figure}[htp]
    \centering
        \includegraphics[width=1\linewidth]
    {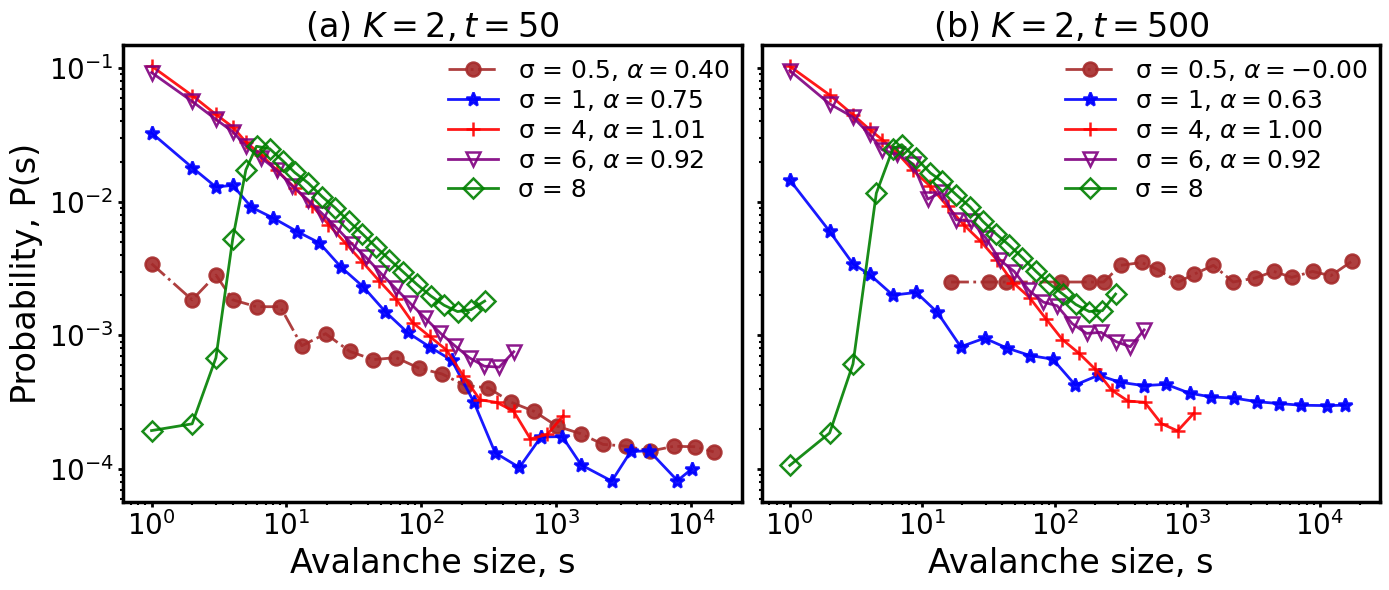}
    \caption{Probability distribution of avalanche sizes for a $999 \times999 $ triangular lattice at $K=2$ of varying disorders($\sigma=0.5,1,2,4,6,8$), obtained by averaging $1000$ localized random field configurations for: a)$t=50$ and b)$t=500$.}
    \label{fig:avalanche_disorder_K2}
\end{figure}

\begin{figure}[htp]
    \centering
        \includegraphics[width=1\linewidth]
        {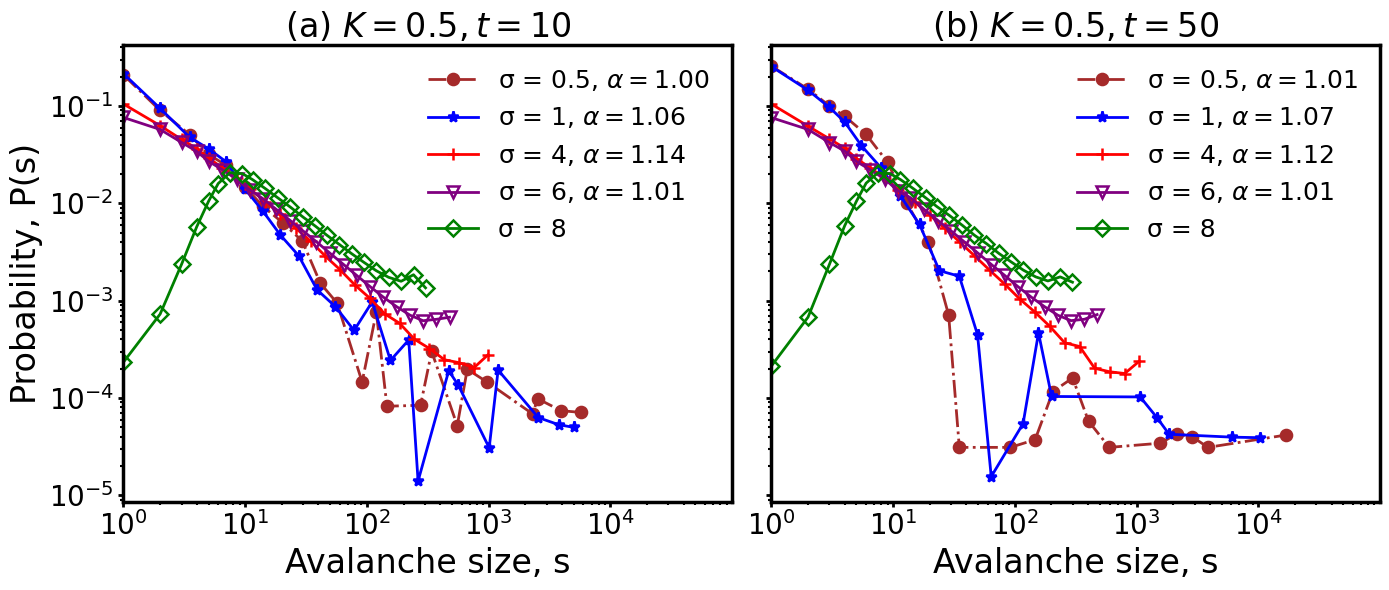}
        \caption{Disorder-dependent avalanche distribution for a $999 \times999 $ triangular lattice at $K=0.5$ with $\sigma=0.5,1,4,6,8$ obtained by averaging $1000$ localized random field configurations for: a)$t=10$ and b)$t=50$.}
    \label{fig:avalanche_disorder}
\end{figure}

In this section, the role of disorder in spin dynamics at low and high temperature is studied. Fig.~\ref{fig:disortrih} shows hysteresis loops on a triangular lattice for disorder $\sigma=0.5, 1, 4, 6, 8$, at two temperature regimes: low ($K=2$) and high ($K=0.5$) for  $t=50$. In both cases, as disorder is increased, the hysteresis loop width (coercivity), remanent magnetization and the slope of magnetization (magnetic susceptibility) decrease. At small $\sigma$, the magnetization exhibits large discrete jumps, which translates to larger range in avalanche size distribution curves as seen in Fig.~\ref{fig:avalanche_disorder_K2} and ~\ref{fig:avalanche_disorder}. 

\vspace{1mm}

We select two values of $t$ from Fig.~\ref{fig:remamdisorder} for each temperature such that the system stays in the non-equilibrium state. Fig.~\ref{fig:avalanche_disorder_K2} shows disorder-dependent avalanche size distributions with $\sigma=0.5, 1, 4, 6, 8$ for $K=2$ (low temperature) at a)$t=50$ and b)$t=500$. Here in fig.~\ref{fig:avalanche_disorder_K2}(a), we observe that the range of the largest avalanche size decreases as $\sigma$ is increased. In other words, the avalanche curves become steeper with a narrowing range as the disorder is increased. This is expected as disorder tends to destroy the correlation between spins. Consequently, we observe the power-law decay exponent $\alpha$ becoming larger on increasing $\sigma$ to a value around $1.14$ and tends to fall at $\sigma=6$ till the power-law disappears.  The curve then develops a pronounced peak in the distribution on $\sigma$ beyond 6. Thus, we conclude that power-law behaviour exists till intermediate disorder and fails beyond $\sigma>6$. 
 This similar trend is observed from Fig.~\ref{fig:avalanche_disorder_K2}(b) when $t=500$. However, with increasing $t$, the distribution progressively flattens.  The initial power-law behaviour in $t=50$ becomes less steep, indicating a higher probability of occurrence of larger avalanches. This trend is expected because increasing $t$ effectively allows the system more time to evolve, facilitating the formation of larger clusters. This flattening is more distinct in smaller disorder $\sigma=0.5, 1, 2$, as the correlation between spins is larger in this case.

For  high temperature $K=0.5$, the avalanche distributions at: a)$t=10$ and b)$t=50$ are shown in {Fig.~\ref{fig:avalanche_disorder}. We chose $t=10$ from Fig.~\ref{fig:remamdisorder} as we observe that in high temperature equilibration is much faster. 
Comparing Fig.~\ref{fig:avalanche_disorder} (a) and (b), we observe that the trend remains qualitatively similar with Fig.~\ref{fig:avalanche_disorder_K2} except that for $K=0.5$, the range of avalanche size is relatively smaller than $K=2$. This is due to larger thermal fluctuations at $K=2$. Additionally, at $t=50$ for $K=0.5$ and $\sigma=6$, the system is already in equilibrium as seen from Fig.~\ref{fig:remamdisorder}, thus it does not follow the power-law behaviour. }
From figures~\ref{fig:disortrih}, \ref{fig:avalanche_disorder_K2}, and \ref{fig:avalanche_disorder}, we can conclude that as the hysteresis loop vanishes, the power-law fails. Power-law behaviour is prevalent from low to intermediate disorder ($\sigma\leq6$) whenever $K$ and $t$ values are within non-equilibrium regime.

\subsection{Varying Temperature}
\label{subsec:VTTL}
\begin{figure}[h!]
\centering
\includegraphics[width=1\linewidth]
{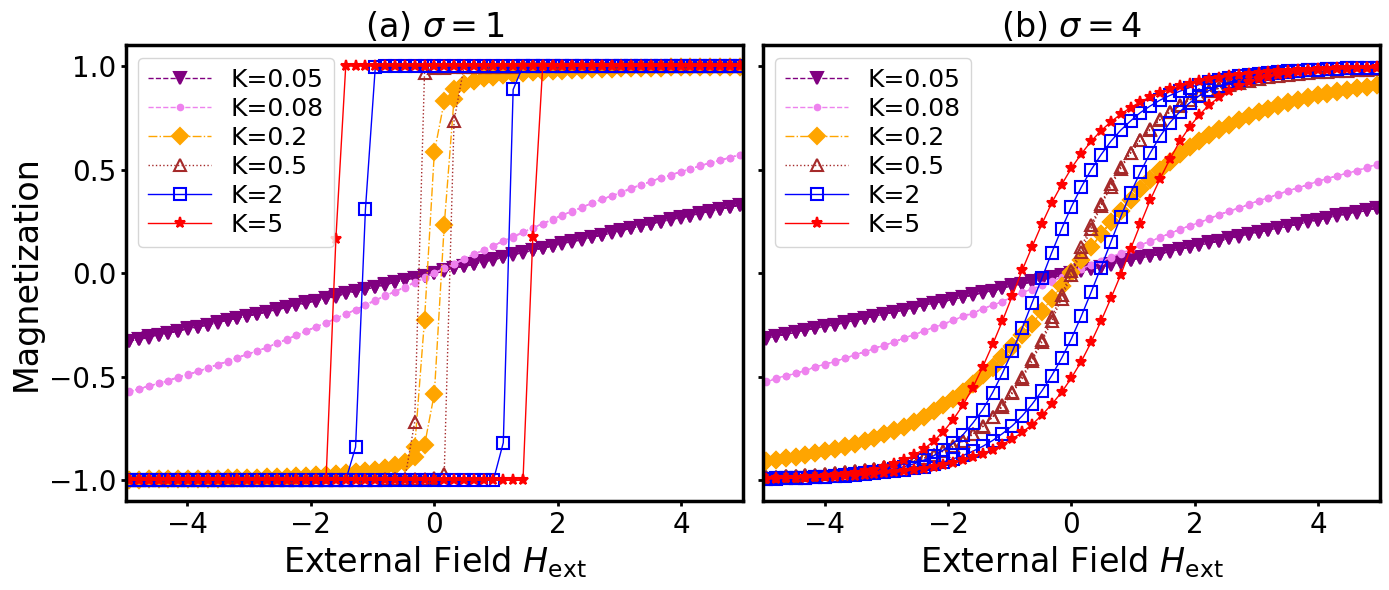}
\caption{Temperature-dependent hysteresis behaviour in a $999\times999$ triangular lattice for a) $\sigma=1$ and b) $\sigma=4$, with $K=0.005, 0.05, 0.08, 0.2, 0.5, 2, 5$ and $t=50$.}
\label{fig:hysttemp}
 \end{figure}

The influence of temperature on a $999\times999$ triangular lattice for fixed disorder is examined through the behaviour of hysteresis curve. Fig.~\ref{fig:hysttemp} shows the hysteresis curve for fixed disorder, $\sigma=1$ and $4$ at $t=50$. As $K$ is increased ($K=1/T$), both the hysteresis loop-width and remanent magnetization increase, suggesting enhanced magnetic retention at lower temperature. It is evident from Fig.~\ref{fig:hysttemp} (a) that when $\sigma=1$ (low disorder), magnetization remains frozen for low temperatures and undergoes an abrupt change once $H_{ext}$ exceeds a critical value, similar to that observed at low-disorder under moderate values of $K$ in the previous section. Fig.~\ref{fig:hysttemp} (b) shows that when $\sigma=4$ (high disorder), there is a smooth rise in the magnetization on increasing $H_{ext}$ as spins flip independently of one another unlike the low disorder case.

\begin{figure}[hbt]
    \centering
        \includegraphics[width=1\linewidth]
   {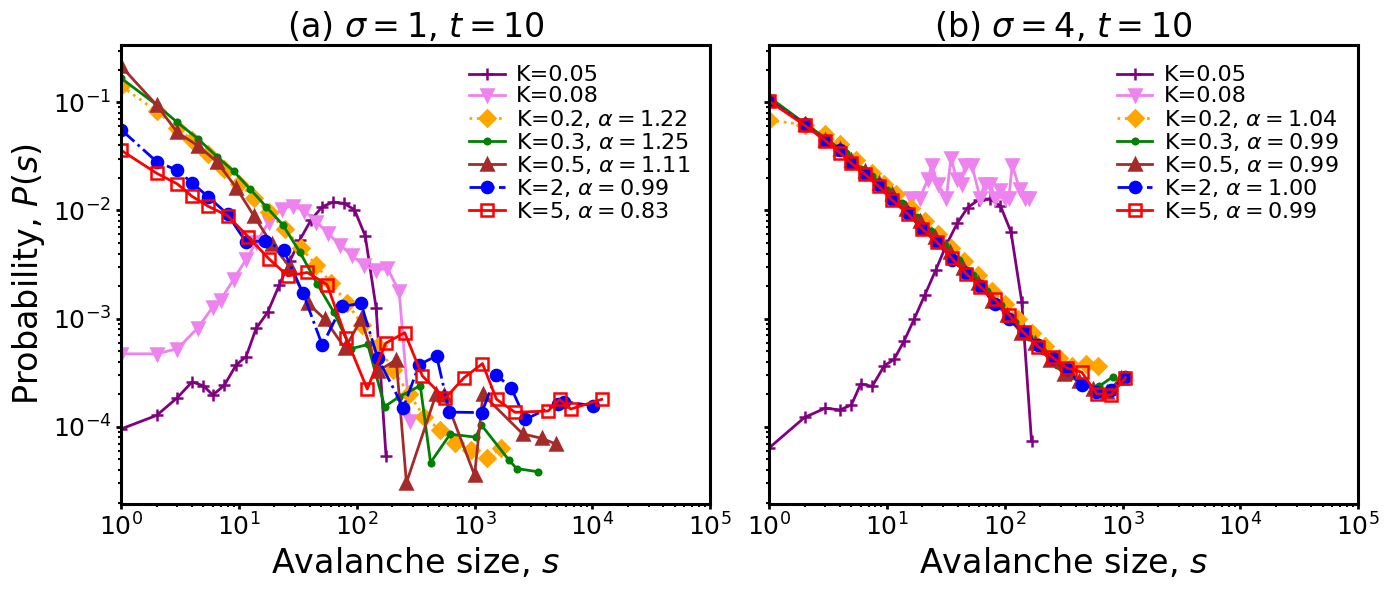}
\caption{Avalanche distribution for a $999 \times999$ triangular lattice of varying temperature ($K=0.005, 0.05, 0.08, 0.2, 0.5, 2, 5$), obtained by averaging $1000$ localized random field configurations at: a)$\sigma=1$ and b)$\sigma=4$. }
    \label{fig:avgavatemp}
\end{figure}

Additionally, we study how temperature affects the distribution of avalanche sizes.  Fig.~\ref{fig:avgavatemp} shows the avalanche size distribution curves with $K=0.05, 0.08, 0.2, 0.3, 0.5, 2, 5$ at fixed disorder, $\sigma=1$ and $4$. We select  $t=10$ from Fig 3 such that the remanent magnetization does not saturate. We observe as the temperature is increased, the range of avalanche sizes decreases. This is expected as large thermal fluctuations randomize the spin orientations, and they flip independently of one another. However, the exponent $\alpha$ increases to a value $1.25$ at $K=0.3$ and then decreases on further increasing $T$ till a pronounced peak appears in the distribution for $K<0.2$ which marks the disappearance of the power-law response for $\sigma=1$.
For large disorder $\sigma=4$, we find $\alpha$ to be around 1 irrespective of $K$ values. Thus, Fig.~\ref{fig:avgavatemp} shows that at low disorder ($\sigma=1$) the region following power-law behaviour exhibits a comparatively stronger temperature dependence than at higher disorder($\sigma=4$). This behaviour is consistent with Barkhausen avalanche statistics in $2D$ Permalloy thin films studied under controlled disorder\cite{Barkhausenexproy}.

Comparing Fig.~\ref{fig:avalanche_disorder} and Fig.~\ref{fig:avgavatemp}, we can conclude that both disorder and temperature have similar effects on the avalanche distribution. When either temperature or disorder is increased, the avalanche curve becomes steeper and the range of the avalanche size decreases gradually before the appearance of a peak.

\section{Conclusion}
This study qualitatively demonstrates the behaviour of spin avalanches at finite temperature on triangular lattices in the framework of RFIM. 
The analysis explores the changes in the hystersis loop and the avalanche distribution in spin system due to interplay of different parameters: relaxation time $t$, disorder $\sigma$, and inverse temperature $K$. These parameters are related to properties of coercivity, remanence and susceptibility of a magnetic material. 

All systems in principle thrive to equilibrium when given an infinite relaxation time. Thus, hysteresis which is a nonequilibrium effect will eventually disappear. However, the rate at which this happens depends on $K$ and $\sigma$ as the spin relaxation probability is governed by these parameters. On performing experiments, how long one can wait for a feasible time becomes important. 
To characterize this equilibration process, we analyze the evolution of remanent magnetization with $t$, as zero remanent magnetization indicates an equilibrium state.
Indeed, we find that remanence decreases on increasing $t$ until it saturates to zero. The route however depends on both temperature and disorder. These findings provide a framework for selecting appropriate values of $t$, enabling the system to be studied either within the equilibrium regime or in the non-equilibrium regime, depending on the requirements of the study.

Once $t$ is fixed, the system is analyzed first by varying $\sigma$ for fixed $K$ and then varying $K$ for fixed $\sigma$. We start with the hysteresis loop and study the associated avalanches.
Analysis of the avalanche size distributions confirms that spin avalanches exhibit power-law behaviour at small disorder and low temperatures, thereby validating earlier theoretical studies conducted on square lattice\cite{finitetemp2d}.
When either temperature or disorder is increased the avalanche curve becomes steeper and the range of avalanche size becomes narrower. 
For a triangular lattice the power-law fails when $\sigma>6$ and $K<0.2$. The region exhibiting power-law behavior shows a comparatively stronger temperature dependence at lower disorder with critical exponent $\alpha$ varying as a function of temperature whereas $\alpha\approx 1$ at high disorder irrespective of $K$. These results agree with the experimental studies on Barkhausen avalanche statistics in 2D Permalloy thin films under controlled disorder.

We extend the investigation to regions beyond the power-law for a broader understanding of the spin avalanches at finite temperature. In this region, the curve develops a pronounced peak in the distribution before decaying exponentially, which is qualitatively similar to  Second Magnetization Peak (SMP) observed on $Al$ doped $YBa_2Cu_3O_{6+\delta}$ single crystals under varying magnetic fields and temperatures\cite{babuflux}. 
We expect that the predicted trends and characteristic curves reported here will provide useful insights in similar experiments.

\section*{Acknowledgements}
DT acknowledges the Department of Science
and Technology, Anusandhan National Research Foundation (ANRF), Government of India, for financial support under POWER Grant No. SPG/2022/000678.
DT also acknowledges partial support from ANRF under the
Partnerships for Accelerated Innovation and Research (PAIR)
for the project JNU-PAIR network on Science for Sustainable
Future, File No. ANRF/PAIR/2025/000029/PAIR.

\bibliography{reference}
\end{document}